\documentclass[12pt]{article}

\usepackage{graphicx} 
\usepackage{amsmath}
\usepackage{subcaption} 
\RequirePackage{xcolor} 
\RequirePackage{amssymb}
\RequirePackage{amsmath} 
\usepackage[macce]{inputenc} 
\RequirePackage[T1]{fontenc}
\RequirePackage{times}
\RequirePackage{mathptmx}
\usepackage[bookmarks,raiselinks,pageanchor,hyperindex,colorlinks,citecolor=black,linkcolor=black,urlcolor=blue,filecolor=black,menucolor=black]{hyperref}

\makeatletter \@addtoreset{equation}{section}
\makeatother 
\renewcommand{\thefootnote}{\alph{footnote}}

\newcommand{\pos}[1]{\href{http://pos.sissa.it/cgi-bin/reader/contribution.cgi?id=#1}{\tt #1}} 
\newcommand{\email}[1]{{\tt\href{mailto:#1}{#1}}} 

\begin{document}
\thispagestyle{empty}
%
 \mbox{} \hspace{1.0cm}
       January 5, 2016
       \mbox{} \hfill BI-TP 2014/26v2\hspace{1.0cm}\\
       \mbox{} \hfill PoS(CPOD2014)070\hspace{1.0cm}\\
       \mbox{} \hfill arXiv: 1501.01127v3 [hep-ph]\hspace{1.0cm}\\
\begin{center}
\vspace*{2cm}
\renewcommand\thefootnote{*}
{{\Large\bf Confining forces \footnote{Poster presentation at the \emph{9th International Workshop on Critical Point and Onset of Deconfinement - CPOD2014}, 17-21 November 2014,
ZiF (Center for Interdisciplinary Research), University of Bielefeld, Germany. \\ First published in \emph{Proceedings of Science}, CPOD 2014 [\pos{PoS(CPOD2014)070}].} \\}}
\addtocounter{footnote}{-1}
\renewcommand\thefootnote{\alph{footnote}}
\vspace*{1.0cm}
{\large Dirk Rollmann$^{1,}$\footnote{E-mail: \email{rollmann@physik.uni-bielefeld.de}} and David E. Miller$^{1,2,}$\footnote{E-mail: \email{dmiller@physik.uni-bielefeld.de} and \email{om0@psu.edu}}
\\}
\vspace*{1.0cm}
${}^1$ {Faculty of Physics, University of Bielefeld, D-33501 Bielefeld, Germany}\\
${}^2$ {Department of Physics, Pennsylvania State University,
Hazleton Campus, \\
Hazleton, PA 18202, USA \\}
\end{center}
\vspace*{2cm} {\large \bf Abstract \\} We discuss the forces on the internal constituents of the hadrons
based on the bag model. The ground state of the hadrons forms a color
singlet so that the effects of the colored internal states are
neutralized. From the breaking of the dilatation and conformal
symmetries under the strong interactions the corresponding currents
are not conserved. These currents give rise to the forces changing
the motion of the internal particles which causes confinement.
 \setcounter{footnote}{0}
 \renewcommand\thefootnote{\arabic{footnote}}
\newpage
\section{Dilatation Current and Conformal Currents}
The starting points of our discussion are the dilatation and the conformal currents \cite{jackiw72}. In four-dimensional Minkowski space $\mathbb{M}^{4}$ the dilatation current is given by
\begin{equation} 
\label{gl01}
D^{\mu}(x)=x_{\nu}T^{\mu\nu}(T)
\end{equation}
and the four conformal currents can be written as
\begin{equation}
\label{gl02}
K^{\mu\alpha}(x)=(2x^{\alpha}x_{\nu}-g^{\alpha}_{\nu}x^2)T^{\mu\nu}(T).
\end{equation}
With the covariant position four-vector $x_{\nu}$ and the inner product of the metric tensor with positiv time metric $g^{\alpha}_{\nu}=g_{\nu\rho} g^{\rho\alpha} = \delta^{\nu}_{\alpha}=\mathbf{1}^{\nu}_{\alpha}$. The four conformal currents are labeled  with $\alpha=0,1,2,3$.
To investigate the conservation of these currents under strong interactions we look at the divergence of these currents
\begin{eqnarray}
\label{gl03}
\partial_{\mu}D^{\mu}(x)&=&T^{\mu}_{\mu}(T)\\
\label{gl04}
\partial_{\mu}K^{\mu\alpha}(x)&=&2x^{\alpha}T^{\mu}_{\mu}(T).
\end{eqnarray}
Thus the divergence relations of these currents  $D^{\mu}(x)$ and $K^{\mu\alpha}(x)$ relate directly to the thermally averaged trace of the energy-momentum tensor $T^{\mu}_{\mu}(T)$.
\section{Breaking of Scale and Conformal Symmetries}
It is well known in quantum chromodynamics (QCD) that the scale and conformal symmetries must be broken to avoid physically absurd mass spectra \cite{jackiw72}. With exact scale and conformal symmetries all particles have to be massless or their mass spectra continuous \cite{pokorski}. Thus the related currents can not be conserved. From the divergence relations in Eqs. (1.3) and (1.4) one can see, that the breaking of scale and conformal symmetries arise from a finite trace of the energy-momentum tensor (\emph{trace anomaly}) \cite{david07}
\begin{equation}
T^{\mu}_{\mu}(T)=\epsilon(T)-3p(T).
\end{equation}
Thus, $T^{\mu}_{\mu}(T)$ acts as an order parameter and gives the magnitude of scale and conformal symmetry breaking.
\section{Lattice Gauge Theorie and Equation of State}
Lattice gauge simulations of QCD compute not directly the trace of the thermally averaged energy-momentum tensor $T^{\mu}_{\mu}(T)$, but the dimensionless \emph{interaction measure} $\Delta(T)$
\begin{equation}
\Delta(T)=(\epsilon(T)-3p(T))/T^4. 
\end{equation}
Here are the differences between the energy density $\epsilon(T)$ and three times the pressure $3p(T)$ strongly suppressed by the division of $T^4$. To obtain the \emph{equation of state} $\epsilon(T)-3p(T)$ as an actual physical quantity one has to multiply the interaction measure by $T^4$
\begin{equation}
\Delta(T) \cdot T^4=\epsilon(T)-3p(T).
\end{equation}
In Ref. \cite{boyd} the numerical values for the interaction measure were calculated from lattice gauge simulations (lattice QCD) for pure SU(3) gauge theory for different lattice sizes. The obtained graphs of the equation of state are shown in Fig. 2 in Ref. \cite{david07}. For all lattice sizes the equation of state shows a rapid growth at the deconfinement temperature $T_{d}$ and a transition to a slower but continual growth in the range where data exists. There is no obvious sign that $T^{\mu}_{\mu}(T)=\epsilon(T)-3p(T)$ decreases to zero at high temperatures. This is also true for the lattice simulations with massive dynamical quarks even if the growth of the equation of state is much slower, see Fig. 9 in Ref. \cite{david07}. That is, the Quark gluon plasma (QGP) is not an ideal ultrarelativistic gas as it would be for $\epsilon(T)=3p(T)$. The thermally averaged trace of the energy-momentum tensor $T^{\mu}_{\mu}(T)$ also enters the gluon condensate \cite{leutwyler}
\begin{equation}  
\langle {G^{2}}\rangle_{T}=\langle{ G^{2}}\rangle_{0} - \langle{T^{\mu}_{\mu}}\rangle_{T}.
\end{equation}
With increasing $T^{\mu}_{\mu}(T)$ the expectation value of the vacuum gluon condensate $\langle{ G^{2}}\rangle_{0}$ is progressively reduced and the gluon condensate $\langle G^{2}\rangle_{T}$ becomes negative and continues to fall. This also holds if Eq. (3.3) is generalized to massive quark fields $m_{q}\langle \overline{\psi_{q}} \psi_{q}\rangle$ \cite{david99}
\begin{equation}
\label{quarkfelder}
\langle G^{2}\rangle_{T} = \langle G^{2}\rangle_{0} + m_{q} \langle \overline{\psi_{q}} \psi_{q}\rangle_{0} - m _{q}\langle \overline{\psi_{q}} \psi_{q} \rangle_{T} -\langle T^{\mu}_{\mu m_{q}}\rangle_{T}.
\end{equation}
For further details see Fig. 10-15 in Ref. \cite{david07}. Since the gluon condensate, as a correlation function of the gluon field strength tensor  $G^{\mu\nu}_{a}$ \cite{leutwyler}
\begin{equation}
G^2 \equiv \frac{-\beta(g)}{2g^3} G^{\mu\nu}_{a}  G_{\mu\nu}^{a},
\end{equation}
does not vanish for all computed temperatures above $T_{d}$, the QGP remains a strongly interacting gas.  Where $a$ denote the color index for $SU(N _{c})$, $\beta(g)$ the renormalization group beta function and $g$ the coupling. However, with a non vanishing trace of the energy-momentum tensor, scale and conformal symmetries under the strong interactions remain broken even with vanishing masses and at high temperatures. This situation is different to the breaking of chiral symmetry, which is restored in the chiral limit $\langle \overline{\psi_{q}} \psi_{q}\rangle_{m_{q}\to 0}$ as well as in the high temperature limit $T \to \infty$.\footnote{For a recent work on chiral symmetry breaking in confined quarkyonic matter see the talk \emph{Inhomogeneous and Quarkyonic phases of High Density QCD} given by L. McLerran at this CPOD 2014 conference [\pos{PoS(CPOD2014)046}].}
 \section{Confining Forces}
The physical insight of the divergence relations in Eqs. (1.3) and (1.4) is breaking of scale and conformal symmetries caused by the trace anomaly. Now we look \cite{david07,david99} at the physical dimension of the corresponding currents in Eqs. (1.1) and (1.2).
The energy-momentum tensor as an energy density is of the dimension of energy per unit volume. This corresponds to a force per area. Multiplied by a length to a square arises a force. Thus
\[
K^{\mu\alpha}(x) {~is~a~force}.
\]
The dilatation current $D^{\mu}(x)$ is of the dimension energy per area or force per length. This motivated the projection of the dilatation current on the coordinate axes: $D^{\mu}(x)x_{\mu}$. Then likewise
\[
D^{\mu}(x)x_{\mu}~is~a~force.
\]
The character of these forces arise from the type of symmetry breaking. The breaking of conformal symmetry give rise to an angular change of the world line of a parton. In a numerical evaluation in two dimensions we will see that $K^{\mu\alpha}(x)$ causes a strong steering effect of the motion of a parton in a confined region. According to \cite{david07} we call $K^{\mu\alpha}(x)$ \emph{fourspan}. The breaking of dilatation symmetry causes a stretching force along the world line of a parton. We call $D^{\mu}(x)x_{\mu}$  \emph{dyxle} \cite{david07}. The relationship between the breaking of dilatation and conformal symmetries and the resulting forces is similar to the relationship between a homogeneous space and momentum conservation. A violation of homogeneity would cause an extra momentum, according to a force.  Only the effects of fourspan and dyxle are different, namely a change of direction.
\section{Ground State and Bag-Model}
The thermally averaged energy-momentum tensor $T^{\mu\nu}(T)$ can be separated into a vacuum part $\theta^{\mu\nu}_{0}$, which is not temperature dependent, and a finite temperature contribution $\theta^{\mu\nu}(T)$ [3]
\begin{equation}
T^{\mu\nu}(T)=\theta^{\mu\nu}_{0}+\theta^{\mu\nu}(T).
\end{equation}
The thermal part $\theta^{\mu\nu}(T)$ is subject to lattice-QCD simulations. For the sake of simplicity we carry out our numerical evaluations in the ground state.
To step around the standard problems with infinities of any ground state we use a bag type of model \cite{chodos} for the vacuum part $\theta^{\mu\nu}_{0}$. In this model the energy density $\epsilon$ in the ground state is given by the \emph{bag-energy} $B$  and the pressure $p$ by $-B$. The bag energy raises the hadron ground state above the QCD vacuum and the negative bag pressure balances the parton pressure to ensure their confinement in a hadron.
Then the trace of the energy-momentum tensor in the ground state in four-dimensional Minkowski space becomes
\begin{equation}
\theta^{\mu}_{\mu0}=\epsilon-3p=4B.
\end{equation}
And in two-dimensional Minkowski space $\mathbb{M}^{2}$ we yield
\begin{equation}
\theta^{\mu}_{\mu0}=2B.
\end{equation}
Then the two conformal breaking forces (now denoted as \emph{twospan}) and the dyxle in two-dimensional Minkowski space, $\mu, \alpha = 0,1$, can be written down as
\begin{equation}
K^{\mu\alpha}(x)=2B
\begin{pmatrix}
(x^{0})^{2}+(x^{1})^{2} & 2x^{1}x^{0}\\
2x^{0}x^{1} & (x^{0})^{2}+(x^{1})^{2}
\end{pmatrix}
\end{equation}
and
\begin{equation}
\label{2ddilatation}
D^{\mu}(x)x_{\mu}=B[(x^{0})^{2}-(x^{1})^{2}]=B\tau^{2}.
\end{equation}
With the time coordinate $x^{0}$ and the spatial coordinate $x^{1}$.
The dyxle $D^{\mu}(x)x_{\mu}$ depends on the proper time squared $\tau^{2}$ corresponding to our interpretation as a stretching force along the world line of a parton.
\section{Transformation on two-dimensional Light-Cone Coordinates}
In two dimensions the conformal matrix Eq. (5.4) is symmetrical and hence only one conformal current is independent. 
Furthermore the two conformal currents are mirrored on the positive light-cone coordinate.
This motivated a transformation on two-dimensional light-cone coordinates
\begin{equation} 
\hat{x}^{\pm}=\frac{1}{\sqrt{2}}(\hat{x}^{0}\pm\hat{x}^{1}).
\end{equation}
In addition we provide the forces with a minus sign corresponding to the confining character of these forces. Then we obtain the two forces of the twospan
\begin{eqnarray}
K^{+}(x^{+})=K^{\mu 0}(x) + K^{\mu 1}(x)=\sqrt{2}\theta^{\mu}_{\mu0} (x^{+})^{2}(-\hat{x}^{+})\\
K^{-}(x^{-})=K^{\mu 0}(x) - K^{\mu 1}(x)=\sqrt{2}\theta^{\mu}_{\mu0} (x^{-})^{2}(-\hat{x}^{-}). 
\end{eqnarray}
After this transformation the $K^{+}(x^{+})$ and $K^{-}(x^{-})$ uncouple and depend only on the  positive $x^{+}$ and negative $x^{-}$ light-cone coordinate, respectively. 
With the transformation
\begin{equation}
x^{+}x^{-}=\frac{1}{\sqrt{2}} (x^{0}+x^{1})\frac{1}{\sqrt{2}}(x^{0}-x^{1})=\frac{1}{2}((x^{0})^{2}-(x^{1})^2)
\end{equation}
we yield for the dyxle
\begin{equation}
D^{\mu}(x)x_{\mu}=-2Bx^{+}x^{-}.
\end{equation}
\section{Numerical Evaluations in 1+1 dimensional Minkowski Space}
\begin{figure}[t] 
	\centering
	\begin{subfigure}[t]{7cm}
	\includegraphics[width=6.5cm]{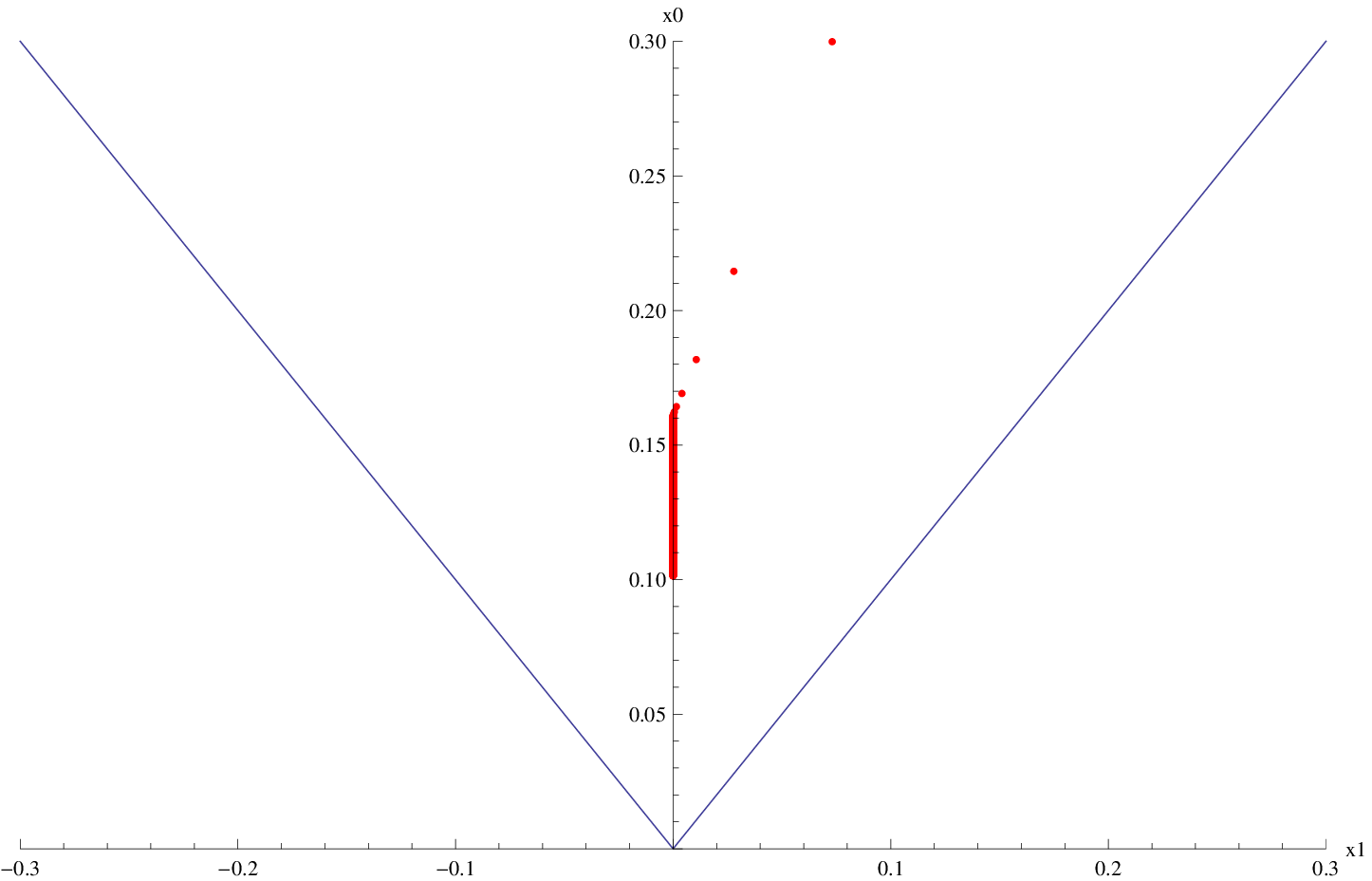} 
	\label{fig:dyxle}
	\end{subfigure}
	\quad
	\begin{subfigure}[t]{7cm}
	\includegraphics[width=6.5cm]{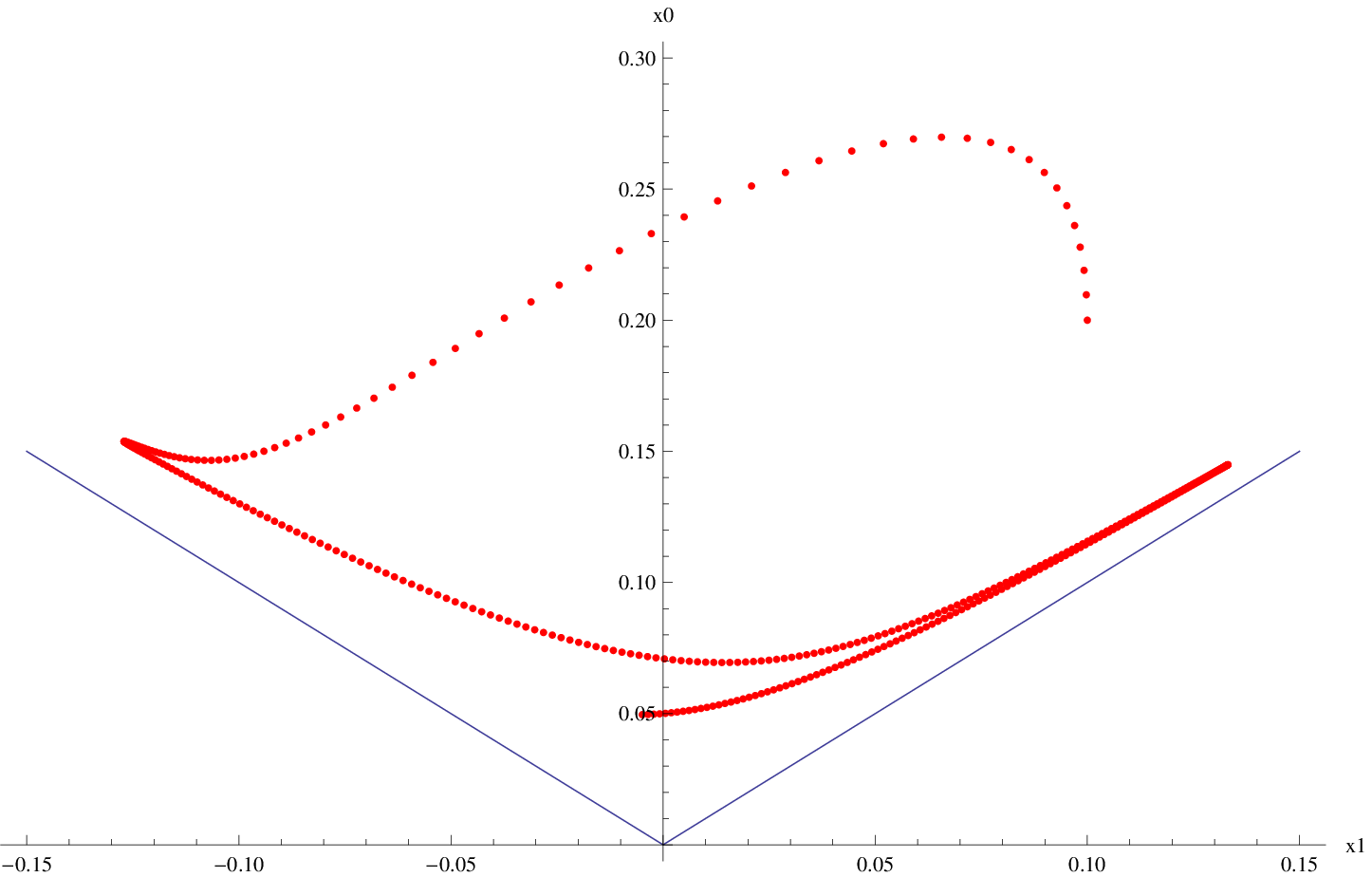}
	\label{subfig:twospan}
	\end{subfigure}
	\caption{(color) Trajectory by the dyxle (left)  with initial values $x^{0}=2.0$, $x^{1}=1.0$, $\Delta t=0.1$, 2000 loops and the twospan (right) with initial values $x^{0}=0.2$, $x^{1}=0.1$, $\Delta t=0.01$, 400 loops.}
\end{figure}
\begin{figure}[b] 
	\centering
	\begin{subfigure}[h]{7cm}
	\includegraphics[width=6.5cm]{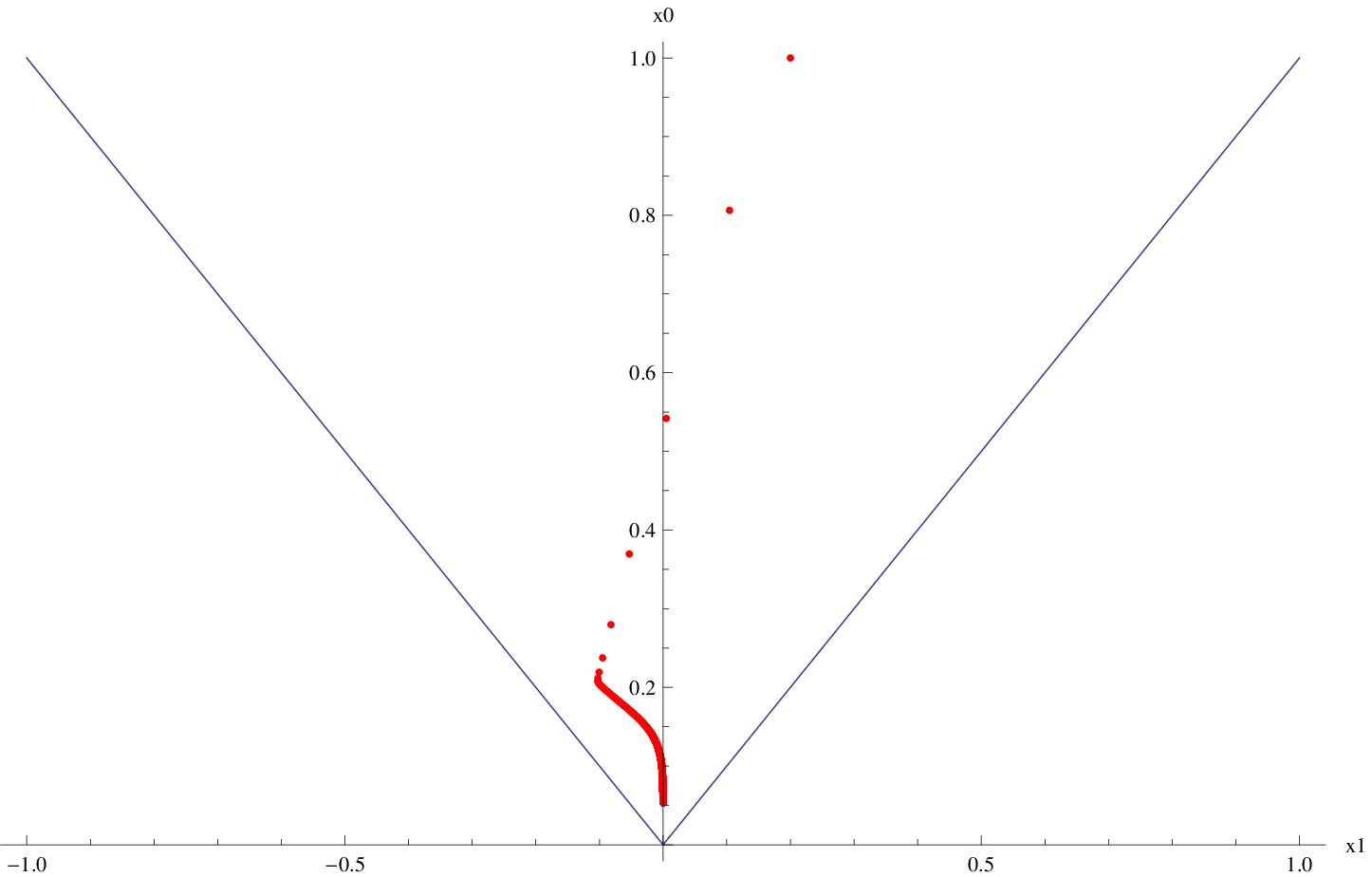} 
	\label{kombiniert2}
	\end{subfigure}
	\quad
	\begin{subfigure}[h]{7cm}
	\includegraphics[width=6.5cm]{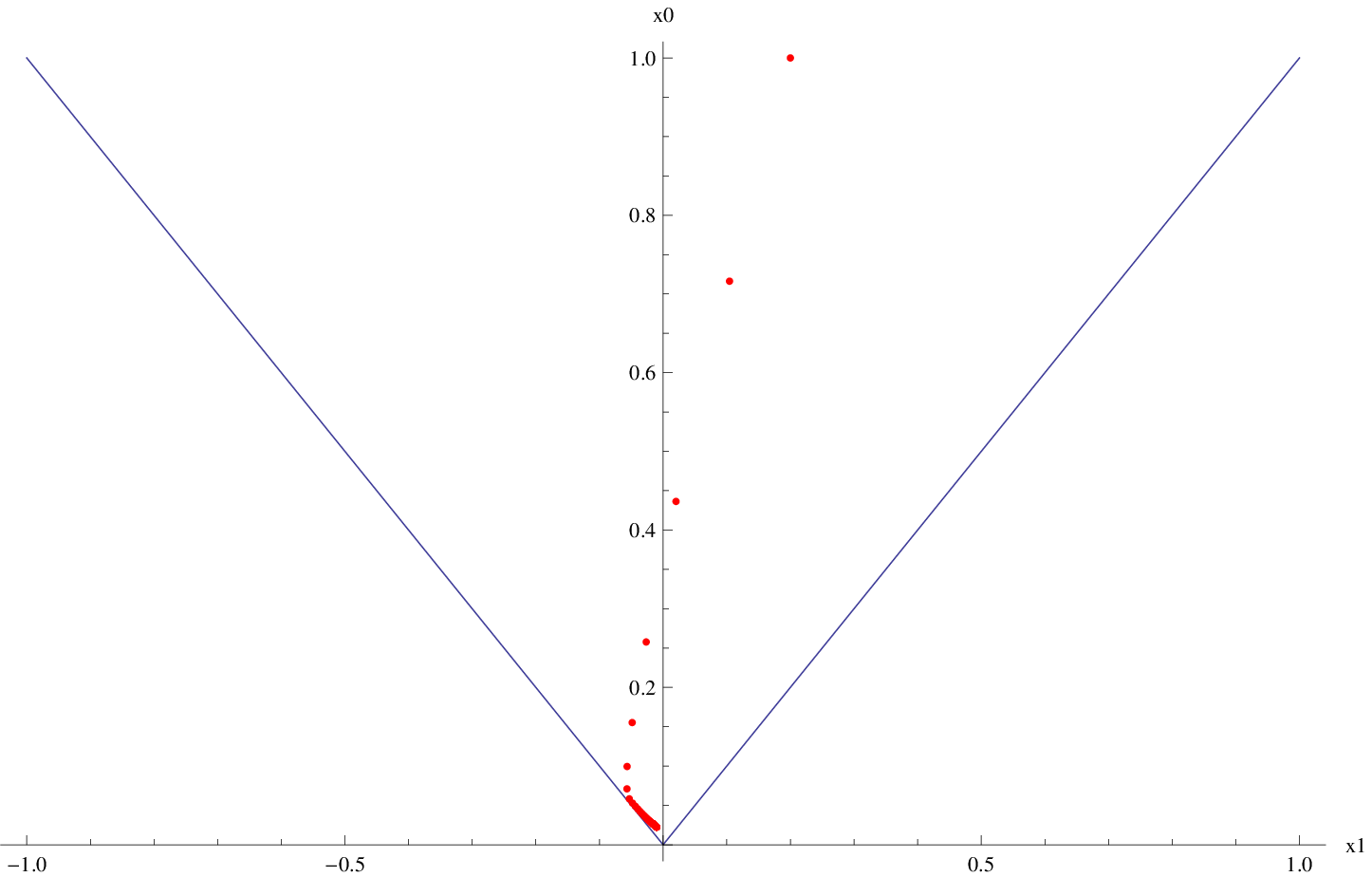}
	\label{kombiniert4}
	\end{subfigure}
	\caption{(color) Trajectories by twospan and dyxle combined. Initial values $x^{0}=1.0$, $x^{1}=0.2$; $\Delta t=0.1$, 10000 loops (left) and with the same parameters, but  smaller time steps $\Delta t=0.01$ (right).}
\end{figure}
In numerical evaluations \cite{dirk} the trajectories of a parton in a bag or hadron which underly twospan and dyxle were calculated in two-dimensional Minkowski space $\mathbb{M}^{2}$ in the ground state. Since the ground state of the hadrons forms a color singlet so that the effects of the colored internal states are neutralized we regard a colorless parton model \cite{feynman69}. We use the law of motion $\vec{F}=m\vec{a}$ and $\vec{v}=\vec{a}\tau$ with the proper time $\tau=\sqrt{(x^{0})^{2}-(x^{1})^{2}}$, since the dyxle itself depends on the proper time. In every loop the time $t$, representing a progressive time, is increased by a time step $\Delta t$. The light-cone coordinates (blue axes) form a natural limit of the bag. The value for the mass is chosen as $m=0.1$ just about the mass of a strange quark $m_{s}=0.095~GeV$. From the vacuum gluon condensate $\langle G^{2}\rangle_{0}=0.012~GeV^4$ \cite{shifman} we chose the bag constant in the ground state as $B=0.01$.
The trajectory of a parton by the dyxle is shown in Fig. 1 (left). The parton moves from the outer light-cone to the origin. Thereby the minus sign in the dyxle, Eq. (5.5), arising from Minkowski metric, causes a subtle directing effect: The parton trajectory is directed to the time axis $x^{0}$. 
In the further course the parton moves toward the origin along the time axis. The changes in the position of the parton are less with decreasing values of $x^{+}x^{-}$ and the proper time $\tau$.
The trajectory of a parton by the twospan is shown on the right side of Fig. 1. $K^{+}(x^{+})$ and  $K^{-}(x^{-})$ act along the positive  $-\hat{x}^{+}$ and negative $-\hat{x}^{-}$ light cone, respectively,  in the direction of the origin. $K^{+}(x^{+})$ vanishes on the negative light cone and $K^{-}(x^{-})$ vanishes on the positive light cone.  At the positive light cone $K^{+}(x^{+})$ takes over and pulls the parton back to the "center" of the bag. In combination of both these forces result a steering effect which pulls the parton from one light cone to the other, approaching the origin. 
The trajectories of a parton by the dyxle and twospan together are shown in Fig. (2). These are the physically relevant evaluations since the trace anomaly breaks both symmetries simultaneously. The parton makes only one movement to the negative light cone and then it moves to the time axis due to the dyxle. Again, the parton first moves from the outer light-cone to the origin and near the origin the changes in the position of the parton are less with decreasing values of $x^{+}x^{-}$ and the proper time $\tau$. The origin acts as fixed point type. In both evaluations in Fig. 2 were calculated 10000 loops of parton displacements. The crucial point of the simulations are the time steps  $\Delta t$. Fig. 2 shows the trajectories for two different time steps  $\Delta t=0.1$ (left) and $\Delta t=0.01$ (right). Smaller time steps could provide more realistic simulations.
\section{Summary and Outlook}
In this numerical evaluation in two-dimensional Minkowski space, twospan and dyxle, the forces which result from the breaking of conformal and scale symmetries in QCD, pull the parton from the boundary to the center of the bag. The forces are particularly strong on the outer light-cone coordinates, which we regard as the limit of a bag or a hadron. We consider this as confinement of a parton in a hadron or more generally as quark confinement. At the origin the forces become particularly weak corresponding to asymptotic freedom of quarks. Concerning the direction of the time we notice that the time beyond the horizon, which is formed by the bag limit, does not have the same meaning like the "normal" physical time.\footnote{To the idea of hadron horizon, see the talk \emph{Hadron Freeze-Out and Unruh Radiation} given by P. Castorina at this CPOD 2014 conference [\pos{PoS(CPOD2014)007}].} But this is still a subject to be investigated. For further statements simulations of the fourspan and the dyxle in four-dimensional Minkowski space are required as well as a refinement of the numerical methods. But we expect that the confining character of the fourspan and the dyxle also will hold in four dimensions. Since the parton couples to the energy, the fourspan and the dyxle could also be significant for deconfined matter and high-energy particle collisions. For instance, the off-diagonal entries of $K^{\mu\alpha}(x)$ correspond to the shearing forces, which also occur in particle collisions.\\

For one of us (DR) the participation at the CPOD 2014 conference was supported by the ExtreMe Matter Institute (EMMI) by Helmholtz Association.

\end{document}